\documentclass[onecolumn]{emulateapj}

\shorttitle{Weibel, Two-Stream, Filamentation, Oblique, Bell, Buneman\ldots}
\shortauthors{Bret}

\begin{document}

\title{Weibel, Two-Stream, Filamentation, Oblique, Bell, Buneman\ldots\\which one grows faster ?}

\author{A. Bret}
\affil{ETSI Industriales, Universidad de Castilla-La Mancha, 13071 Ciudad Real, Spain}

\begin{abstract}
Many competing linear instabilities are likely to occur in astrophysical settings, and it is important to assess which one grows faster for a given situation. An analytical model including the main beam plasma instabilities is developed. The full 3D dielectric tensor is thus explained for a cold relativistic electron beam passing through a cold plasma, accounting for a guiding magnetic field, a return electronic current and moving protons. Considering any orientations of the wave vector allows to retrieve the most unstable mode for any  parameters set. An unified description of the Filamentation (Weibel), Two-Stream, Buneman, Bell instabilities (and more) is thus provided, allowing for the exact determination of their hierarchy in terms of the system parameters. For relevance to both real situations and PIC simulations, the electron-to-proton mass ratio is treated as a parameter, and numerical calculations are conducted with two different values, namely 1/1836 and 1/100. In the system parameters phase space, the shape of the domains governed by each kind of instability is far from being trivial. For low density beams, the ultra-magnetized regime tends to be governed by either the Two-Stream or the Buneman instabilities. For beam densities equalling the plasma one, up to four kinds of modes are likely to play a role, depending of the beam Lorentz factor. In some regions of the system parameters phase space, the dominant mode may vary with the electron-to-proton mass ratio. Application is made to Solar Flares, Intergalactic Streams and Relativistic shocks physics.
\end{abstract}



\section{Introduction}
Weibel, Filamentation, Two-Stream, Bell or Buneman instabilities are ubiquitous in astrophysics. They are involved in the physics of Solar Flares where relativistic electron beams are assumed to lose their energy through beam-plasma instabilities, producing hard X-ray emissions \citep{Karlicky,KarlickyApJ}. Filamentation, or Weibel, instabilities could also be responsible for the birth of cosmological magnetic fields \citep{Schlickeiser2003,schlickeiser2005,Lazar2008,Lazar2009} as unstable particle streams through the intergalactic medium can magnetize an initially un-magnetized system. Such instabilities could also play an important role in explaining the origin of a variety of high energy photons sources including Supernova Remnants, Active Galactic Nuclei, Gamma Ray Bursts or Pulsar Wind Nebulae \citep{Piran1999,Gedalin2002,Piran2004,Waxman2006,StockemApJ}. Within some of these systems, it is assumed that cosmic rays are accelerated through shocks (relativistic or not) while the instability generated upstream by their interaction with the interstellar medium provides the magnetic turbulence eventually responsible for synchrotron radiation emissions \citep{Medvedev1999,SilvaApJ,nishikawa,FrederiksenApJ2004,Milos2006,Lemoine2006,Niemec2008}.

Regardless of the context, the typical structure investigated consists in a beam-plasma system initially both charge and current neutralized. Initial charge neutralization implies the inclusion of positive and negative species, while current neutralization demands at least two streaming species. Streams are often considered as opposed, but they can be parallel when streaming species are of opposite signs. For example, a pair beam does not need anymore beams to be current neutral. Finally, accounting for an external magnetic field $\mathbf{B}_0$ allows the system to be relevant to a wide class of astrophysical problem. The simplest case consists in a $\mathbf{B}_0$ parallel to the flow(s) but normal or oblique orientations have also been considered \citep{Pelletier1991,DieckmannApJ,BretPoPOblique,SiloniApJ}.

We consider here the generic system formed by a cold electron beam of density $n_b$ streaming at initial velocity $\mathbf{v}_b$, with Lorentz factor $\gamma_b=(1-v_b^2/c^2)^{-1/2}$, over a cold electron/proton plasma. The beam current is neutralized by an electronic return current of density $n_p$ and velocity $\mathbf{v}_p$ such that $n_b\mathbf{v}_b=n_p\mathbf{v}_p$. Ions are initially at rest with density $n_i=n_b+n_p$. Finally, the model includes a flow-aligned magnetic field $\mathbf{B}_0\parallel \mathbf{v}_b$. Even for such a simple system, linear stability analysis is intricate because unstable modes are numerous. The intent of this paper is to clarify this issue and determine the fastest growing mode for any given set of parameters, by implementing an exact model encompassing every basic instability. In order to keep the results tractable, emphasis is made on the simplest possible model with the lowest number of free parameters. As explained in the sequel, these requirements demand a beam-plasma system with mobile ions and a guiding magnetic field, while kinetic effects will not be investigated here. A cold fluid model is thus exactly solved, rendering every possible coupling between unstable modes. The dispersion equation arising from the exact dielectric tensor is  analyzed without calling on the electrostatic ($\mathbf{k}\parallel \mathbf{E}$), or the purely electromagnetic ($\mathbf{k}\perp \mathbf{E}$), approximations. Such kind of calculation is mandatory if the Two-Stream or Buneman electrostatic modes are to be described within the very same formalism than the electromagnetic Filamentation or Bell-like instabilities.

Let us list the possible unstable modes by progressively ``assembling'' the system. The first block here is the un-magnetized background plasma with fixed protons. Because we take it cold, it is stable against every kind of perturbations. Note that some temperature anisotropy would make it Weibel unstable \citep{Weibel}, the fastest growing modes being found for wave vectors perpendicular to the high temperature axis \citep{Kalman1968}. Kinetic effects can thus drive this first component unstable even before we add anything else.

Let us now ``add'' the electron beam and let the plasma electrons establish the return-current (protons are still fixed). To simplify the discussion, we consider here a diluted beam with $n_b\ll n_p$ (the rest of the paper also deals with higher beam densities). As is known, the resulting beam-plasma system is unstable. Perturbations with wave vector $\mathbf{k}\parallel\mathbf{v}_b$ are unstable for $0<k\lesssim\omega_p/v_b$ and define the so-called Two-Stream Instability \citep{BhomGross}. Perturbations having $\mathbf{k}\perp\mathbf{v}_b$ are also unstable for any $k$ and pertain to the Filamentation instability\footnote{The Filamentation instability is singled out from the Weibel one. In its original context, the Weibel instability results from a temperature anisotropy with no drift. Filamentation instability involves relative streaming of various species. In the present context, both instabilities maybe be disconnected as one Filamentation stable, or unstable, beam can interact with a Weibel stable, or unstable, plasma \citep{BretPRE2004,BretPRE2005}. They can also interfere with each other \citep{BretPoPFilaWeibel,LazarFilaWeibel,Lazar2008,Lazar2009}.}. Finally, perturbations with $\mathbf{k}$ neither parallel nor perpendicular to the flow are also unstable \citep{fainberg,califano1,califano2,califano3} and may even be the fastest growing modes in the diluted relativistic beam regime \citep{BretPoPHierarchie,BretPRL2008}.

Let us now give the protons the possibility to move. Every aforementioned unstable modes are still unstable as the two counter-streaming electron beams still interact. But both electron currents can now interact with the protons, giving rise to Buneman unstable modes \citep{Buneman}. A closer look at the situation shows that Buneman modes arising from the beam/protons interaction merge with the  beam/return-current modes  \citep{BretPoPIons}. But the \emph{return}-current/proton interaction results in unstable Buneman modes reaching their maximum growth rate for $k\sim (n_p/n_b)\omega_p/v_b$ and $\mathbf{k}\parallel\mathbf{v}_b$. Within the diluted beam regime, $n_b\ll n_p$ implies unstable modes with much shorter wave length than the Two-Stream modes. Indeed, it has been found than the Buneman modes can compete with the Two-Stream ones if the beam is relativistic enough.

We finally make our system complete by adding a flow-aligned magnetic field. Here again, the new ingredient brings in more unstable modes without necessarily stabilizing the previous ones. To start with, electronic Cyclotron and Upper-Hybrid modes are destabilized, adding new branches to the unstable spectrum for any orientations of $\mathbf{k}$ \citep{Godfrey1975}. Also, Alfv\'{e}n modes resulting from the combination of the vibrating protons in the magnetic field can also be destabilized. This unstable modes where first pointed out by Bell by means of a MHD formalism \citep{Bell2004,Bell2005}, and their description was later extended through the kinetic one \citep{ZweibelApJ,Reville2006}.

Our simple magnetized beam/plasma system with moving protons in eventually Two-Stream, Filamentation, Oblique, Cyclotron, Upper-Hybrid, Buneman, Bell\ldots unstable! As the system is released from equilibrium, every instabilities are triggered, and the outcome of the linear phase is mostly shaped by the growth of the fastest growing one. This most unstable mode can be found for an oblique wave vector so that the search of the dominant mode requires the implementation of a model capable of describing any unstable mode for any  orientation of the wave vector. Note that recent works involving ion beams also evidenced such kind of modes \citep{Niemec2008,Ohira2008}. We thus now proceed to the elaboration of the simplest possible model incorporating all the aforementioned instabilities.

The unstable modes described here are just too numerous to be detailed one by one. On the other hand, each one needs to be discussed since the present aim is precisely to determine their hierarchy. The full wave vector dependance of each mode is therefore skipped, the focus being set on the most unstable wave vector with the corresponding growth rate. Additionally, modes which have not been found to govern the system for any given sets of parameters are only briefly discussed in order to keep the presentation tractable. Still for clarity, the names of the numerous modes have not been abbreviated. Finally, analytic expressions in terms of the magnetic field parameter $\Omega_B$ defined by Eqs. (\ref{eq:variables}) are essentially derived for $\Omega_B>1$, while results are also presented for the un-magnetized regime.

The paper is structured as follow: the analytic model is explained in the next section. Unstable modes found for flow-aligned wave vectors are listed in Section \ref{sec:kpara} before an overview of the 2D unstable spectrum is given in Section \ref{sec:2D}. The key results of the article are exposed in Section \ref{sec:hierar} where the fastest growing mode is determined in terms of the system parameters. In this respect, Figures \ref{fig:4} and \ref{fig:hierar-alf1} can be considered as the main results of the papers. They show which kind of mode governs the linear phase of the beam-plasma system in terms of the parameters. Application is then made to various astrophysical settings in Section \ref{sec:app}, before the final discussion and conclusion.

\section{Analytic model}\label{sec:model}
The model relies on the cold relativistic fluid equations for the three species involved. The calculation follows the lines of previous ones \citep{califano3,Kazimura} except that protons are allowed to move while a static flow-aligned magnetic field is accounted for. The basic equations include Maxwell's equations and,
\begin{equation}\label{eq:consev}
    \frac{\partial n_j}{\partial t}+\nabla\cdot(n_j\mathbf{v}_j)=0,
\end{equation}
\begin{equation}\label{eq:euler}
    \frac{\partial \mathbf{p}_j}{\partial t}+(\mathbf{v}_j\cdot\nabla)\mathbf{p}_j=q_j\left[\mathbf{E}+\frac{\mathbf{v}_j\times(\mathbf{B}+\mathbf{B}_0)}{c}\right],
\end{equation}
where $j=b$ for the electron beam, $j=p$ for the plasma return-current and $j=i$ for plasma ions (protons). Here, $\mathbf{p}_j=\gamma_jm_j\mathbf{v}_j$ and $q_j$ are the momentum and electric charge of specie $j$.

Although lengthy, the derivation of the dielectric tensor is quite standard. The conservation, Maxwell's and Euler's equations are first linearized assuming every quantity slightly departs from  equilibrium  like $\exp(\imath\mathbf{k}\cdot \mathbf{r}-\imath\omega t)$ where $\imath^2=-1$. Since $\mathbf{B}_0\parallel\mathbf{v}_b,\mathbf{v}_p$ while $\sum q_jn_j=\sum q_jn_j\mathbf{v}_j=0$, the charge and current neutral equilibrium state considered exactly fulfills the full set of equations. We can write $\mathbf{k}=(k_x,0,k_z)$ without loss of generality  by virtue of the axial symmetry with respect to the beam and magnetic field axis \citep{Godfrey1975}. We choose the $z$ axis for the direction of the beam and of the magnetic field, having therefore $\mathbf{v}_b=(0,0,v_b)$, $\mathbf{v}_p=(0,0,v_p)$ and $\mathbf{B}_0=(0,0,B_0)$. The linearized conservation and Euler's equations are first used to express the perturbed total current $\mathbf{J}_1$ in terms of the electromagnetic field. The first order magnetic field is then eliminated through $\mathbf{B}_1=(c/\omega)\mathbf{k}\times \mathbf{E}_1$ and the resulting expression of the current is inserted into the usual combination of Maxwell-Faraday and Maxwell-Amp\`{e}re equations, namely
\begin{equation}\label{eq:Maxwell}
    \frac{c^2}{\omega^2}\mathbf{k}\times(\mathbf{k}\times \mathbf{E}_1)+\mathbf{E}_1+\frac{4 \imath \pi}{\omega}\mathbf{J}_1(\mathbf{E}_1)=0.
\end{equation}
The electrostatic ($\mathbf{k}\times\mathbf{E}_1=0$) or purely electromagnetic approximations ($\mathbf{k}\cdot\mathbf{E}_1=0$) would here result in a simplification of the $\mathbf{k}\times(\mathbf{k}\times \mathbf{E}_1)$ term. Although such options may be useful when focusing on one give mode,  the risk at this stage would be to loose informations about possible dominant modes which would not fit in. A \emph{Mathematica} Notebook has been implemented to  symbolically compute the tensor \citep{BretCPC}. It is expressed in terms of the dimensionless variables,
\begin{equation}\label{eq:variables}
  x = \frac{\omega}{\omega_p}, ~~\mathbf{Z}=\frac{\mathbf{k} v_b}{\omega_p},~~ \alpha=\frac{n_b}{n_p},~~
  \beta=\frac{v_b}{c},~~\Omega_B = \frac{\omega_b}{\omega_p},~~R=\frac{m_e}{m_i},
\end{equation}
where $\omega_p^2=4\pi n_p q^2/m_e$ is the electronic background plasma frequency, $\omega_b=|q|B_0/mc$ the non-relativistic electronic cyclotron frequency and $m_e$, $m_i$ the electron and ion (proton) mass respectively. The Alfv\'{e}n velocity $v_A$ can be expressed in terms of the variables above as,
\begin{equation}\label{eq:va}
    \frac{v_A}{c}=\sqrt{R (1+\alpha )} \Omega_B,~~\mathrm{with}~~v_A=\frac{B_0}{\sqrt{4\pi n_i m_i}}.
\end{equation}
The dielectric tensor has the form,
\begin{equation}\label{eq:tensor}
  \mathcal{T}=\left(%
\begin{array}{ccc}
  T_{xx}    &  T_{xy}    &   T_{xz}  \\
  T_{xy}^*  &  T_{yy}    &   T_{yz}  \\
  T_{xz}^*  &  T_{yz}^*  &   T_{zz}  \\
\end{array}%
\right),
\end{equation}
where $z^*$ is the complex conjugate of $z$. Tensor elements are reported in Appendix \ref{app:tensor} in terms of the dimensionless variables (\ref{eq:variables}). In the limit of zero magnetic field, only the square of the charges appears in the tensor expression. As a consequence, the results in such case also apply to two counter-streaming \emph{pair plasmas} \citep{Jaroschek2005,RamirezRuiz2007}.

The most general expression of the dispersion equation then reads $\det\mathcal{T}=0$ and necessarily encompasses every possible unstable modes. As previously mentioned, we here emphasize the search of the most unstable mode $\mathbf{k}_M$ of the unstable $\mathbf{k}$ spectrum. The function $\mathbf{k}_M$ can only depend on the beam-to-plasma density ratio $\alpha$, the beam Lorentz factor $\gamma_b$, the magnetic field strength parameter $\Omega_B$ and the mass ratio $R$. Once the former has been fixed to 1/1836, or any higher value more suited to comparison with PIC simulations \citep{Jaroschek2004,Spitkovsky2008,KarlickyApJ}, the most unstable mode can only be a function of $(\alpha, \gamma_b,\Omega_B)\in [0,1]\times[1,\infty]\times[0,\infty]$. Note that this study is not limited to the diluted beam case so that the return-current flowing at $v_p=\alpha v_b$ can also reach relativistic velocities. Tables \ref{tab:diluted} and \ref{tab:nondiluted} summarize the results of the two next sections as they display the main unstable modes in the diluted ($\alpha\ll 1$) and symmetric ($\alpha=1$) regimes respectively. Modes properties in the magnetized regime are given only for $\Omega_B>1$. A detailed study (at $R=0$) of the transition from $\Omega_B=0$ can be found in \cite{Godfrey1975}.

\begin{table*}
\caption{\label{tab:diluted}Main unstable modes in the diluted beam regime.}
\begin{ruledtabular}
\begin{tabular}{llll}
Modes                   &     $Z_x$ &       $Z_z$\footnotemark[*]     &    Growth rate \\
\hline
$T_{xx} - \imath T_{xy}$ &     0                            &  $\Omega_B/\gamma_b$  &  $\propto\sqrt{\alpha}R\Omega_B$\footnotemark[**]\\
$T_{xx}+ \imath T_{xy}$ &     0                            &  $\frac{1}{2}\alpha/\Omega_B$  &  $\propto\alpha R\Omega_B$\footnotemark[**]\\
Two-Stream             &     0                            &  1           &  $\frac{\sqrt{3}}{2^{4/3}}\alpha^{1/3}/\gamma_b$\\
Buneman                &     0                            &  $1/\alpha$  &  $\frac{\sqrt{3}}{2^{4/3}}R^{1/3}$\\
\hline
Oblique$_{B0}$, $\Omega_B=0$                  &     $\infty$                       &  $1$  &  $\frac{\sqrt{3}}{2^{4/3}}(\alpha/\gamma_b)^{1/3}$\\
UHL\footnotemark[***], $\Omega_B>1$      &     $\infty$                       &  $\Omega_B/\gamma_b$  &  $\frac{1}{2}(\alpha/\Omega_B)^{1/2}$\\
Oblique, $\Omega_B>1$                &     $\sqrt{1+\Omega_B^2}$           &  $\sqrt{1+\Omega_B^2}$  &  $\frac{1}{4}\alpha^{1/3}/\gamma_b$\\
UHL\footnotemark[***], $\Omega_B>1$      &     $\infty$                       &  $\Omega_B/\gamma_b+\sqrt{1+\Omega_B^2}$  &  $\frac{1}{2}\alpha^{1/2}/\Omega_B$
\end{tabular}
\end{ruledtabular}
\footnotetext[*]{For a given $Z_x$, the growth rate is a function of $Z_z$. The table only mentions the $Z_z$ and the growth rate pertaining to the fastest growing mode. Magnetized mode are reported for $\Omega_B>1$. The transition between $\Omega_B=0$ and 1 is described in \cite{Godfrey1975} for $R=0$.}
\footnotetext[**]{Bounded by Eq. (\ref{eq:bound}).}
\footnotetext[***]{UHL stands for ``Upper-Hybrid-Like''.}
\end{table*}

\begin{table*}
\caption{\label{tab:nondiluted}Main unstable modes in the high density beam regime $\alpha=1$.}
\begin{ruledtabular}
\begin{tabular}{llll}
Modes                   &   $Z_x$ &       $Z_z$     &    Growth rate \\
\hline
$T_{xx} \pm \imath T_{xy}$ &     0     &    & $\lesssim\sqrt{R}$ \\
TSB\footnotemark[*], $\gamma_bR^{1/3}< 1/2$             &     0                 &  $\sqrt{3}/2\gamma_b^{3/2}$         &  $1/2\gamma_b^{3/2}$\\
TSB\footnotemark[*], $\gamma_bR^{1/3}>1/2$             &     0                            &  $\sqrt{2R}$         &  $\frac{\sqrt{3}}{2^{7/6}}R^{1/6}/\gamma_b$\\
\hline
Oblique, $\Omega_B>1$                    &     $2\Omega_B^{1/2}/\gamma_b^{0.75}$   &  $\Omega_B/2\gamma_b$ &  $0.5/\gamma_b^{1.15}$  \\
UHL\footnotemark[**] 1\& 2, $\Omega_B>1$                 &     $\infty$                    &  $\sim\Omega_B/\gamma_b$  &  $1/\Omega_B$  \\
\end{tabular}
\end{ruledtabular}
\footnotetext[*]{TSB stands for the merged Two-Stream/Buneman mode.}
\footnotetext[**]{UHL stands for ``Upper-Hybrid-Like''.}
\end{table*}

\section{Flow-aligned wave vectors}\label{sec:kpara}
We here detail the case of flow-aligned wave vector instabilities before analyzing the most general situation. Even if these modes do not always govern the system, such analysis is useful to get familiar with the competing instabilities, and make the bridge with previous investigations. When considering $\mathbf{k}=(0,0,k_z)$, the tensor (\ref{eq:tensor}) simplifies to,
\begin{equation}\label{eq:tensorzz}
  \mathcal{T}=\left(%
\begin{array}{ccc}
  T_{xx}    &  T_{xy}    &   0       \\
  -T_{xy}  &  T_{yy}    &   0       \\
  0         &  0         &   T_{zz}  \\
\end{array}%
\right),
\end{equation}
so that the dispersion equation simply reads,
\begin{equation}\label{eq:disperzz}
   T_{zz}(T_{xx}+\imath T_{xy})(T_{xx}-\imath T_{xy})=0.
\end{equation}
This  equation already contains most of the unstable modes already discussed. The $T_{zz}$ generates electrostatic Two-Stream and Buneman modes while the others are responsible for the so-called ``Bell'' unstable electromagnetic modes.

\subsection{Electrostatic modes}
The $T_{zz}$ terms reads,
\begin{equation}\label{eq:Tzz}
    T_{zz}=1-\frac{R(1+\alpha)}{x^2}-\frac{\alpha }{(x-Z_z)^2 \gamma_b^3}-\frac{1}{(x+\alpha Z_z)^2\gamma_p^3}.
\end{equation}
The proton contribution stems from the term $\propto R/x^2$, the beam one from the second term with numerator $\alpha$, and the return-current with  $\gamma_p=(1-v_p^2/c^2)^{-1/2}$ accounts for the last factor. The dispersion equation for these modes is independent of the magnetic field since the resulting electrostatic modes have $\mathbf{k}\parallel \mathbf{E}$ and only generate velocity perturbations along the magnetic field. This equation has been analyzed elsewhere \citep{BretPoPIons}. For a diluted beam, the interaction of the two electron beams produces Two-Stream unstable modes with maximum growth rate \citep{Mikhailovskii1}
\begin{equation}\label{eq:TS}
    \delta=\frac{\sqrt{3}}{2^{4/3}}\frac{\alpha^{1/3}}{\gamma_b}~~\mathrm{for}~~Z_z=1,
\end{equation}
and the interaction of the return-current with the proton yields unstable Buneman modes at much smaller wavelength with,
\begin{equation}\label{eq:Bun}
    \delta=\frac{\sqrt{3}}{2^{4/3}}R^{1/3}~~\mathrm{for}~~Z_z=1/\alpha.
\end{equation}
These expressions change when the density ratio approaches unity because the beam and the return-current become symmetric for $\alpha=1$ (see Sec. \ref{sec:diluted}).

\begin{figure}
\begin{center}
 \includegraphics[width=0.4\textwidth]{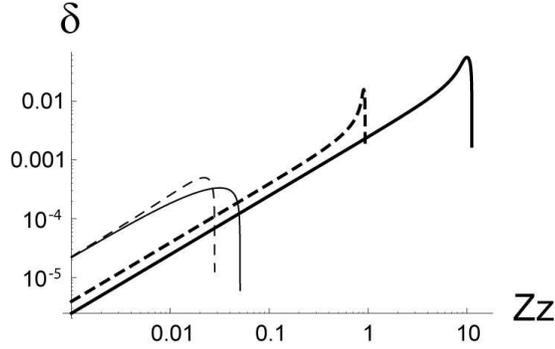}
\end{center}
\caption{Growth rate $\delta$ ($\omega_p$ units) of the four unstable modes for flow-aligned wave vectors in terms of $Z_z=k_\parallel v_b/\omega_p$. Parameters are $\alpha=0.1$, $\gamma_b=20$, $\Omega_B=1$ and $R=1/1836$. The two electromagnetic modes yield the thin plain and dashed curves from Eq. (\ref{eq:Bellzz}) with ``-'' and ``+'' respectively. The bold dashed pertains to the Two-Stream modes and the Buneman modes gives the plain bold one.}
\label{fig:1}
\end{figure}

\subsection{Electromagnetic modes}
The unstable electromagnetic modes arise here from the $T_{xx}\pm \imath T_{xy}$ terms of Eq. (\ref{eq:disperzz}) which read,
\begin{equation}\label{eq:Bellzz}
   T_{xx}\pm \imath T_{xy}=x^2-\frac{Z_z^2}{\beta^2}+
   \frac{\alpha (Z_z-x)}{ \gamma _b(x-Z_z )\mp \Omega_B}
   -\frac{x+\alpha Z_z}{\gamma _p(x+\alpha Z_z)\mp\Omega_B}
   -\frac{R x (1+\alpha )}{x\pm R \Omega_B}.
\end{equation}
A straightforward calculation from the general dispersion relation for circularly polarized wave with such wave vectors \citep{Ichimaru,Achterberg1983,Amato2009} yields the very same result. These modes are the present configuration analogs to the ones explained by Bell   \citep{Bell2004,Bell2005}. The main difference with the scenarios in which they are usually involved is that they are presently driven by an \emph{electronic} current rather than by a baryonic one.

Both dispersions equations yield here unstable modes. The most relevant feature with respect to the mode hierarchy is that growth rates are bounded by $R\Omega_B$, regardless of the parameters involved. This does not imply the growth rate can go to infinity because unstable solutions exist only in a limited region of the phase space ($\alpha,\gamma_b,\Omega_B,R$).

The $T_{xx}- \imath T_{xy}$ term can be partially assessed analytically neglecting the $x^2$ and the return-current factor. The most unstable modes are found at
$Z_z\sim\Omega_B/\gamma_b$, with $x\sim R\Omega_B$. The maximum growth rate varies like $\sqrt{\alpha}$, and is always bounded by
\begin{equation}\label{eq:bound}
\delta < R\Omega_B-\Omega_B^2R^{3/2}.
\end{equation}
The $T_{xx}+ \imath T_{xy}$ factor is more involved and has been partially treated numerically. Unstable modes are here found at $Z_z\sim\frac{1}{2}\alpha/\Omega_B$, with maximum growth rate varying like $\alpha$, and still bounded by Eq. (\ref{eq:bound}). Note that these modes are stable for $\gamma_b\lesssim\alpha\Omega_B^2$ so that instability in the diluted beam regime demands a low Lorentz factor.

Figure \ref{fig:1} displays the growth rate of these four unstable modes in terms of $Z_z$ for some parameters triggering them all. One can observe how the relevant spectrum extends over 3 orders of magnitude. This point, and its consequences, is discussed in the conclusion.

\begin{figure}
\begin{center}
 \includegraphics[width=0.4\textwidth]{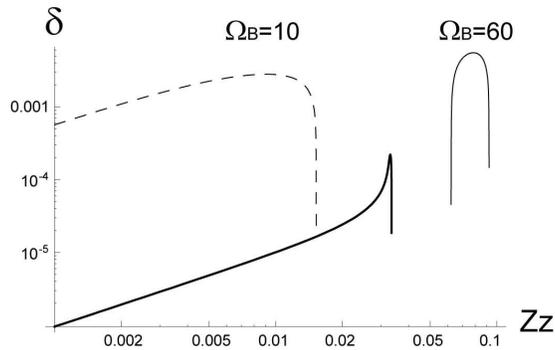}
\end{center}
\caption{Growth rate $\delta$ ($\omega_p$ units) for flow-aligned wave vectors in terms of $Z_z=k_\parallel v_b/\omega_p$. Parameters are $\alpha=1$ for $\Omega_B=10$ and 60, with $R=1/1836$ and $\gamma_b=10^3$. Here, electromagnetic modes (thin lines) can overcome the electrostatic ones (bold line).}
\label{fig:3}
\end{figure}

\subsection{Non-diluted beam regime}\label{sec:diluted}
Because we study small perturbations of the system composed by both the beam \emph{and} the plasma, the present theory is perfectly valid up to $\alpha=1$. However, the physical interpretation of the calculations can no longer be derived considering the beam is a perturbation to the isolated background plasma. The system beam+plasma has its own proper modes, which are quite different from the ones of the isolated plasma. Taking $R=0$ for example, the dispersion equation has two branches exactly given by,
\begin{equation}\label{eq:branches_sym}
    x^2=Z_z^2+\frac{1-\sqrt{1\pm4 Z_z^2 \gamma_b^3}}{\gamma_b^3},
\end{equation}
which definitely differs from the cold isolated plasma dispersion relation $x=1$, i.e $\omega=\omega_p$. For such symmetric system, the Buneman modes at $Z_z\sim 1/\alpha$ merge with the Two-Stream modes at $Z_z\sim 1$, and depending on $\gamma_b R^{1/3}$, the resulting mode may be governed by the electrons or the ions dynamic (see Table \ref{tab:nondiluted}).

For parallel wave vectors, electrostatic modes have been studied previously \citep{BretPoPIons}, and two different regimes need to be considered depending on the product $\gamma_bR^{1/3}$. For $\gamma_b R^{1/3}\lesssim 1/2$, the fastest growing mode has
\begin{equation}\label{eq:BunTSAlpha1NR}
    \delta=\frac{1}{2\gamma_b^{3/2}}~~\mathrm{for}~~Z_z=\frac{\sqrt{3}}{2\gamma_b^{3/2}}.
\end{equation}
Note that this result is exact for $R=0$. In the opposite limit $\gamma_b R^{1/3}\gtrsim 1/2$, one has
\begin{equation}\label{eq:BunTSAlpha1R}
    \delta=\frac{\sqrt{3}R^{1/6}}{2^{7/6}\gamma_b}~~\mathrm{for}~~Z_z=\sqrt{2R}.
\end{equation}

\begin{figure*}
\begin{center}
 \includegraphics[width=0.95\textwidth]{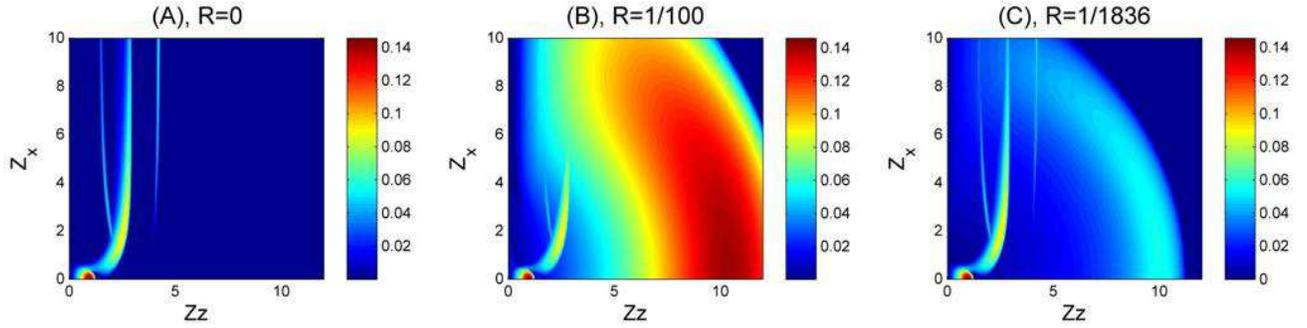}
\end{center}
\caption{(Color online) 2D unstable spectrum in terms of the reduced wave vector $\mathbf{Z}$ for $\alpha=0.1$, $\gamma_b=2$ and $\Omega_B=3$. (A) $R=0$. (B) $R=1/100$. (C) $R=1/1836$.}
\label{fig:2}
\end{figure*}

Regarding the electromagnetic modes, the two terms $T_{xx}\pm \imath T_{xy}$ in Eq. (\ref{eq:disperzz}) yield the same growth rate because for $\alpha=1$, these two functions are equal through the transformation $x\rightarrow -x$. Here again the growth rate of these modes is always bounded by $R\Omega_B$, but instability occurs only for $\gamma_b\lesssim1/R$ and $\Omega_B\lesssim\sqrt{\gamma_b}$. As a consequence, the most unstable mode for a given couple ($\gamma_b,\Omega_B$) cannot grow faster than $R\Omega_B<R\sqrt{\gamma_b}<\sqrt{R}$.

Figure \ref{fig:3} displays the growth rate of the two modes just discussed for parameter sets yielding a Bell governed system. Note that such situation demands a highly magnetized system. Although this plot gives the sensation that Bell-like modes can dominate some region of the parameters phase space, they were never found to govern the full 2D spectrum. This example illustrates the importance of accounting for the full unstable $\mathbf{k}$ spectrum instead of focusing on a given wave vector orientation. While Bell-like modes grow slower for non flow-aligned wave vectors,  other modes grow faster in oblique directions. As a result, the unstable spectrum ends up governed by the latters, even if the formers can ``reign'' over the beam axis. This situation basically stems from the necessary smallness of the mass ratio $R$. Dealing with a proton beam would allow to reach $R=1$, where Bell modes should definitely be found governing some portion of the phase space. A similar theory accounting for a proton beam would be needed to describe such situations.

\section{An overview of the 2D spectrum}\label{sec:2D}
Before we establish the full hierarchy of the unstable modes in terms of the parameters $(\alpha,\gamma_b,\Omega_B)$, let us give an overview of the unstable spectrum in terms of $\mathbf{k}=(k_x,0,k_z)$. The basic principle here is that the unstable modes previously discussed for $k_x=0$ remain unstable for some $k_x\neq 0$. The analytical analysis becomes more involved because all modes are coupled, as the tensor elements $T_{xz}$ and $T_{yz}$ in Eq. (\ref{eq:tensor}) no longer vanish. However, the list of unstable modes is easier to grasp in the diluted beam regime where the beam is just a perturbation to the isolated background plasma. We therefore single out the diluted and non-diluted beam regime in our analysis. Furthermore, unstable modes found in oblique direction  have finite or infinite $k_x$. Both kinds of modes can be tackled by different approaches, and are thus studied separately.

\subsection{Diluted beam - $Z_x=\infty$}\label{sec:2D_diluted}
For the un-magnetized system with infinite ion mass, the resulting spectrum is now well understood, including within the framework of a full kinetic relativistic theory \citep{BretPRL2008}. There is a continuum of unstable modes bridging the Two-Stream instability ($\mathbf{k}_\perp =0$) with the Filamentation instability ($\mathbf{k}_\parallel =0$). The magnetized version  of the same system brings two additional oblique branches. For $\Omega_B>1$, the first one is found at $Z_z \sim \Omega_B/\gamma_b$, with a maximum growth rate varying like $\frac{1}{2}\sqrt{\alpha/\Omega_B}$. Still for $\Omega_B>1$, the second branch is located at $Z_z \sim\Omega_B/\gamma_b+\sqrt{1+\Omega_B^2}$ with a maximum growth rate varying like $\frac{1}{2}\sqrt{\alpha}/\Omega_B$ \citep{Godfrey1975}. Within the same region of the spectrum, the oblique modes already present at $\Omega_B=0$ near $Z_z=1$ are now found at $Z_z\sim\sqrt{1+\Omega_B^2}$ with a maximum increment varying like $\alpha^{1/3}/\gamma_b$. These later modes are commented in Sec. \ref{sec:ObliFinite}. At this stage, the unstable spectrum may look like the one pictured on Fig. \ref{fig:2}(A) for the parameters specified in caption.

\begin{figure*}
\begin{center}
 \includegraphics[width=\textwidth]{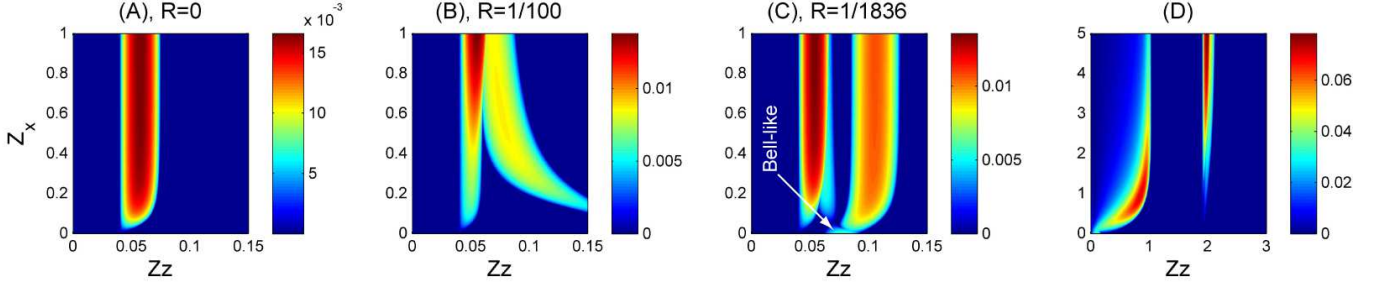}
\end{center}
\caption{(Color online) 2D unstable spectrum in terms of the reduced wave vector $\mathbf{Z}$ for $\alpha=1$, $\gamma_b=10^3$ and $\Omega_B=60$. (A) $R=0$. (B) $R=1/100$. (C) $R=1/1836$. With the same parameters than Fig. \ref{fig:3}, we see here how Bell-like modes around $Z_z=0.08$ do govern the $Z_z$ axis but not the all spectrum. Map (D), with parameters $\Omega_B=10$, $\gamma_b=5$ and $R=1/1836$, clearly evidences the oblique mode explained in Sec. \ref{sec:ObliFinite} for $Z_z,Z_x\sim 1$.}
\label{fig:33}
\end{figure*}

At this junction, some comment is needed to clarify the meaning of ``oblique'' mode (with small ``o''). Figures \ref{fig:2} \& \ref{fig:33} make it clear that unstable modes with both $k_x\neq 0$ and $k_z\neq 0$ are numerous. Such kind of modes have been so far refereed to in the literature as ``Oblique'' \citep{Watson,Niemec2008,Ohira2008}, ``electromagnetic beam-plasma instability'' \citep{califano3}, ``coupled Two-Stream Weibel'' \citep{Jaroschek2005} or ``Mixed mode'' \citep{frederiksenPoP2008}. The problem with such labeling  is that ``oblique modes'' are here just too numerous for only one tag. In Tables \ref{tab:diluted} \& \ref{tab:nondiluted}, as in the rest of the paper, ``oblique modes'' bridging between Two-Stream and  Filamentation instabilities for $\Omega_B=0$ are labeled ``Oblique$_{B0}$''. For $\Omega_B>1$, they evolve into what we presently call ``Oblique'' modes (capital ``O''), with both $k_x\neq 0$ and $k_z\neq 0$, and a growth rate scaling like $\gamma_b^{-1}$. Finally, modes found reaching their maximum growth rate at finite $Z_z$ and $Z_x=\infty$ are labeled ``Upper-Hybrid-Like'' modes after \cite{Godfrey1975}.

The spectrum is even richer when ions are ``allowed'' to move, as Buneman and Bell-like modes are triggered. The resulting spectrum is pictured on Figs. \ref{fig:2}(B,C) for $R=1/100$ and $1/1836$ respectively. An electron-to-proton mass ratio $R=1/100$ may be relevant for PIC simulation where the mass ratio is usually incremented from its 1/1836 realistic value in order to speed up the ions dynamic, saving thus computer time. Regarding this last point, it is interesting to note that such a trick is possible because the growth rates associated with the moving ions increase with $R$, while the related scale length does not (see Eqs. \ref{eq:Bun}). This is reflected on Figs. \ref{fig:2}(B,C) where the dominant modes do not migrate with varying $R$. If such was not the case, some higher mass ratio could demand higher spatial resolution so that the computing time lost by a finer spatial resolution would not necessarily be compensated by the accelerated dynamic. Also on these Figures, the occurrence of the Buneman instability around $Z_z\sim 1/\alpha=10$ is clear. The Two-Stream instability governs here the system even for $R=1/100$, but the electron and ion spectrums are almost disconnected form each other only for $R=1/1836$. Because of the smallness of the real mass ratio, both the time and length scales of each spectrum are disconnected from each other. In such case, the overall spectrum is nearly the superposition of each sub-spectrum.

\subsection{Non-diluted beam regime, $\alpha=1$ and $Z_x=\infty$}\label{sec:2D_Ndiluted}
Without moving ions ($R=0$), the most relevant features of the unstable spectrum in this case are oblique resonances at $Z_z\sim\Omega_B/\gamma_b$ and $Z_z\sim\Omega_B/2\gamma_b$. Only the first one extends up to $Z_x=\infty$ and yields a maximum growth rate $\delta_m\sim 1/\Omega_B$ \citep{BretPoPMagne}. The next paragraph is devoted to an overview of the second kind of modes. This large $Z_x$ regime can be analytically investigated in the following way. We start deriving the dispersion equation in this limit setting $\alpha=1$, $\gamma_b=\gamma_p$ and $Z_x\rightarrow\infty$ in Eqs. (\ref{eq:tensor}). We then use the fact that unstable modes are found with $\Re(\omega)\sim 0$ and develop the resulting dispersion equation to the second order in $x$. In this large $Z_x$ limit, the system is found unstable only between,
\begin{equation}\label{eq:Zinf}
    Z_{z1}=\frac{\Omega_B}{\gamma_b}\sqrt{1-\frac{2 \beta^2 \gamma_b}{\Omega_B^2}},
\end{equation}
and
\begin{equation}\label{eq:Zsup}
    Z_{z2}=\frac{\Omega_B}{\gamma_b}.
\end{equation}
These results are exact and imply that Filamentation instability is canceled as soon as $Z_{z1}>0$, i.e. $\Omega_B>\beta\sqrt{2\gamma_b}$.  For $\Omega_B\gg\beta\gamma_b^{1/2}$, the unstable range decreases like $Z_{z2}-Z_{z1}\sim\beta^2/\Omega_B$. The range of unstable wave vectors becomes therefore increasingly narrow, yielding a quasi-monochromatic unstable spectrum in this limit.

When ion motion is accounted for, part of the stability domain can still be calculated exactly following the same guidelines. With $R\neq 0$, one finds
\begin{equation}\label{eq:ZinfR}
    Z_{z1,R}=Z_{z1},
\end{equation}
and
\begin{equation}\label{eq:ZsupR}
    Z_{z2,R}=\frac{\Omega_B}{\gamma_b}
    \sqrt{\frac{1+R (\gamma_b+\Omega_B^2/2)}{1+R \Omega_B^2 /2}}.
\end{equation}
The maximum growth rate within this range remains close to $1/\Omega_B$, and another range of unstable wave vectors and similar growth rate appears at slightly larger $Z_z$'s. These new modes, arising purely from the ion motion, will not be investigated here although they are accounted for in the forthcoming numerical evaluation of modes hierarchy.

Figures \ref{fig:33}(A-C) shows the 2D spectrum for the parameters chosen for Fig. \ref{fig:3} with $\Omega_B=60$. The purpose of this choice is to observe how some oblique modes govern the full spectrum while Bell-like modes dominate the $Z_z$ axis. On Fig. \ref{fig:3}, the spectrum for $\Omega_B=60$ is clearly governed by these modes at $Z_z\sim 0.08$. The same modes, together with the way they evolve when leaving the $Z_z$ axis, are perfectly visible on Fig. \ref{fig:33}(C) and one can check how they quickly stabilize for oblique wave vectors.

\subsection{Oblique modes at finite $Z_x$}\label{sec:ObliFinite}
Sections \ref{sec:2D_diluted} \& \ref{sec:2D_Ndiluted} discussed unstable modes located at $Z_x=\infty$. The cold fluid approximation typically send some fast growing modes to $Z_x=\infty$, whereas temperature effects tend to stabilize these small wavelengths instabilities by preventing the pinching of too small filaments \citep{Silva2002}. But the occurrence of a magnetic field also triggers some truly locally most unstable oblique modes. The term ``locally'' means here that the very same mode, as defined by one root of the dispersion equation, reaches its maximum for a finite $Z_x$. For more clarity, Figure \ref{fig:33}(D) displays a 2D spectrum with $\alpha=1$, while the other parameters have been chosen to single out this mode.

For the diluted beam case, Table \ref{tab:diluted} mentions a maximum growth rate varying like $\alpha^{1/3}/\gamma_b$. For the symmetric case $\alpha=1$, Table \ref{tab:nondiluted} indicates a maximum oblique growth rate $\sim 0.5/\gamma_b^{1.15}$ (numerical fit). Such scaling of this finite $Z_x$ mode bears importance consequences regarding modes hierarchy: a look at Tables \ref{tab:diluted} \& \ref{tab:nondiluted} shows that the ultra-magnetized regime tends to stabilize every modes, except this one, the Buneman and the Two-Stream. While more modes compete for moderate $\Omega_B$, the large $\Omega_B$ limit is decided among these three. More details on the ultra-magnetized regime are given in Section \ref{sec:ultraBdiluted}.

\section{Modes hierarchy}\label{sec:hierar}
We finally turn to the determination of the modes hierarchy in terms of the parameters $\alpha$, $\gamma_b$ and $\Omega_B$. Only the case $\Omega_B=R=0$ has been treated so far  \citep{BretPoPHierarchie,BretPRL2008} evidencing two main features. On the one hand, Oblique$_{B0}$ modes can govern the diluted beam regime from $n_b/n_p\sim 0.53$. For higher density ratios, Filamentation may dominate depending on the beam Lorentz factor $\gamma_b$. On the other hand, Oblique$_{B0}$ modes govern the ultra-relativistic regime unless the system in strictly symmetric with $n_b=n_p$.

After having calculated symbolically the dielectric tensor using a previously designed \emph{Mathematica} Notebook, the polynomial dispersion equation (17th degree) has been transferred to \emph{MatLab} for  numerical study. A systematic search of the most unstable modes has thus been conducted in terms of $\alpha$, $\gamma_b$ and $\Omega_B$ for $R=1/1836$ and 1/100. For each parameter set ($\alpha,\gamma_b,\Omega_B$), the program returns the $Z_z$ and $Z_x$ components of the most unstable mode together with the real and imaginary parts of the corresponding complex frequency $\omega_r+\imath\delta$.

\begin{figure*}
\begin{center}
\includegraphics[width=\textwidth]{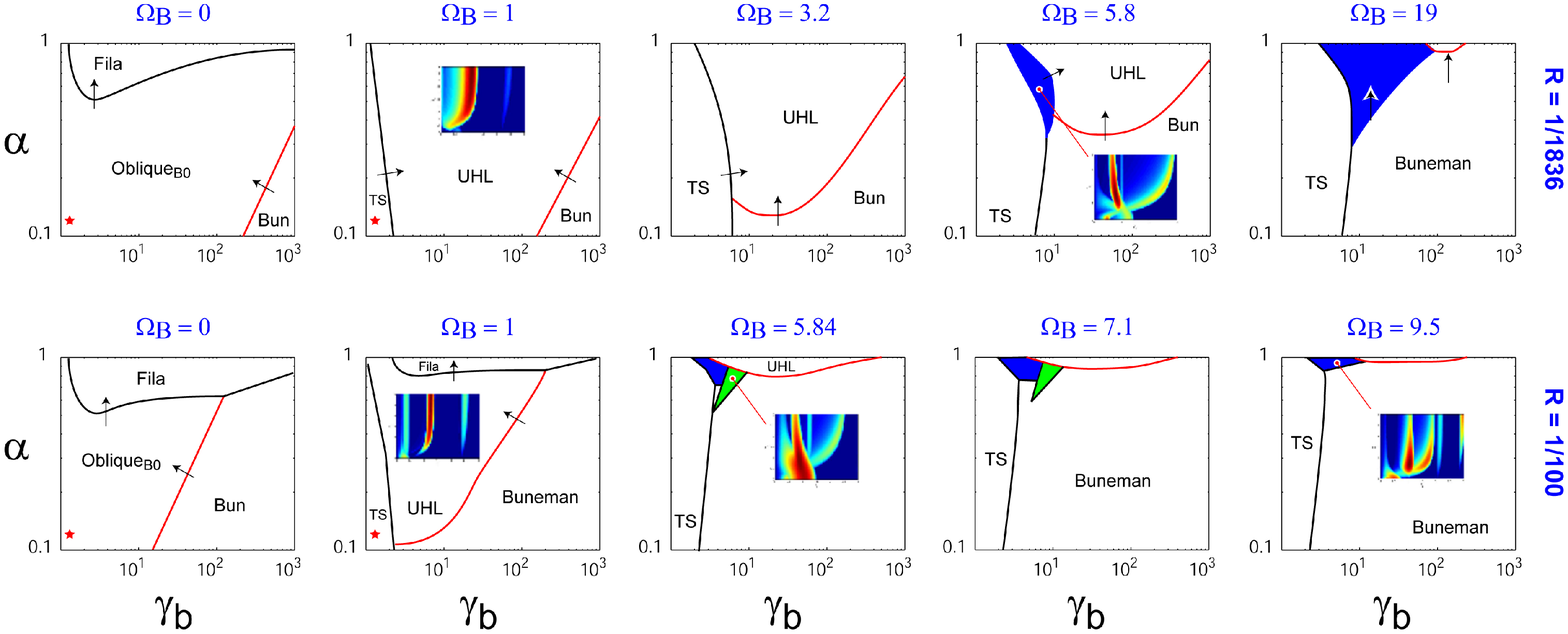}
\end{center}
\caption{Frontiers between the domains governed by different modes in terms of $\alpha$ and $\gamma_b$, for various $\Omega_B$ and $R=1/1836$, 1/100. The arrows on the frontiers show how they evolve when $\Omega_B$ increases. The red crosses show the parameter sets for the Solar Flares application explained in Sec. \ref{sec:flares}.}
\label{fig:4}
\end{figure*}

The function $\delta_M=\max\{\delta, (k_x,k_z)\in \mathbb{R}^2\}$ is clearly continuous. But the real part $\omega_M$ corresponding to $\delta_M$, together with the wave vector $\mathbf{Z}_M$ defining the mode growing at $\delta_M$, may well evolve discontinuously as the most unstable mode can perfectly ``jump'' from one location of the 2D spectrum to another \citep{BretPRL2008}. For example, the $Z_z$ component of the fastest growing mode along the beam axis, switches abruptly from $Z_z=1$ to $1/\alpha$ when the Buneman mode overcomes the Two-Stream one. Indeed, these discontinuities are a way to determine the domains where any given mode governs the system: as long as the parameters evolve in such a way that the very same mode keeps dominating, $\mathbf{Z}_M$ and $\omega_M$ are continuous\footnote{The roots of a polynomial are continuous functions of its coefficients. See \cite{Uherka1977} for a proof.}. But when the dominant mode changes, one of these functions will suffer some discontinuity. Within the present cold model, the most unstable oblique modes are often ``sent'' to $Z_z=\infty$ so that an oblique-oblique transition will not necessarily trigger some discontinuity of the perpendicular component. But this is neither the case of the parallel component $Z_z$ nor of the real part of the frequency $\omega_M$.  Figure \ref{fig:4} has thus been elaborated exploiting these properties and is commented in the sequel. Here again, and for better clarity, we single out the diluted beam regime $\alpha\ll 1$ from the symmetric case $\alpha=1$. The main reason for such a structure is that only these two extremes are  analytically accessible, whereas the intermediate case definitely requires numerical assistance.

\subsection{Diluted beam}
Gathering the aforementioned data, the frontiers between the domains governed by different modes are sketched in terms of $(\gamma_b,\alpha)$ on Fig. \ref{fig:4}. Although $\alpha$ is limited to the range $[0.1,1]$, the lower part of the graph can be easily deduced from the diluted beam regime expressions reported in Table \ref{tab:diluted}. This process is repeated for various $\Omega_B$'s and $R=1/1836$, 1/100.

Starting with $\Omega_B$ =0,   the findings of \cite{BretPoPIons} are here confirmed numerically. Filamentation  dominates for large beam-to-plasma density ratios. The frontier visible in the lower-right corner pertains to the Oblique/Buneman transition. The growth rate of the un-magnetized oblique competing mode in this case\footnote{The ``Oblique$_{B0}$'' of Table \ref{tab:diluted}.} reads $\sqrt{3}/2^{4/3}(\alpha/\gamma_b)^{1/3}$ \citep{fainberg}. The equation of the border is therefore simply,
 \begin{equation}\label{eq:Bun-OBli}
    \alpha=R\gamma_b.
 \end{equation}
For $\alpha=R$ and $\gamma_b=1$ , Eqs. (\ref{eq:TS},\ref{eq:Bun}) are identical so that this frontier extends all the way down to $\alpha=R$. Bellow this line, Buneman modes govern the linear phase.

\begin{figure}
\begin{center}
 \includegraphics[width=0.45\textwidth]{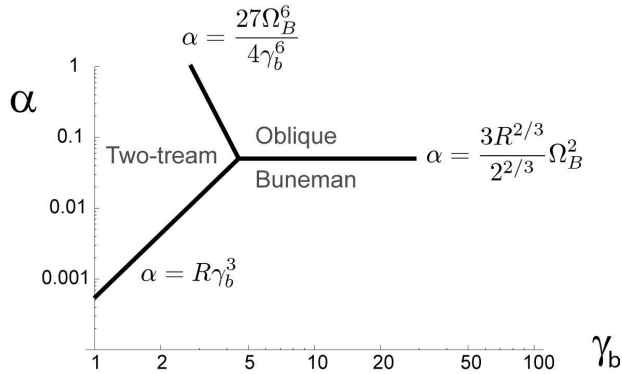}
\end{center}
\caption{Low $\alpha$ partition of the phase space defined by Eqs. (\ref{eq:TS-Obli},\ref{eq:Ob-Bun},\ref{eq:Bun-TS}) for $1\lesssim\Omega_B\lesssim R^{-1/3}$  (see Sec. \ref{sec:ultraBdiluted}).}
\label{fig:partition}
\end{figure}

From $\Omega_B=1$, the evolution is three-fold. To start with, the Two-Stream instability governs an increasing weakly relativistic region (lower-left corner) as its magnetized growth rate remains the same, while oblique modes are made less unstable. By virtue of the same kind of effects, Buneman modes gain weight on the lower-right corner because their growth rate too does not vary with the magnetic field. Finally, Filamentation domain shrinks as the magnetic field progressively shuts it down \citep{Cary1981,StockemApJ}. Regarding the border of the Two-Stream region, a look at the $Z_x$ component shows it switches from 0 to $\infty$ when crossing the limit. A comparison between the growth rate values in this region and the expressions gathered in Table \ref{tab:diluted}, shows that the Two-Stream mode competes with the Upper-Hybrid-Like one located at $Z_z\sim\Omega_B/\gamma_b+\sqrt{1+\Omega_B^2}$. The Two-Stream/Oblique frontier is thus defined for $\alpha\ll 1$ by,
\begin{equation}\label{eq:TS-Obli}
  \frac{\sqrt{3}}{2^{4/3}}\frac{\alpha^{1/3}}{\gamma_b}=\frac{\sqrt{\alpha }}{2 \Omega_B}
  \Leftrightarrow \alpha=\frac{27 \Omega_B^6}{4 \gamma_b^6}.
\end{equation}
Regarding the Buneman/Upper-Hybrid-Like frontier, the growth rate $\sim \frac{1}{2}\sqrt{\alpha}/\Omega_b$ simply decreases with $\alpha$ until the Buneman mode overcomes it. This transition occurs thus for,
\begin{equation}\label{eq:Ob-Bun}
    \frac{\sqrt{3}}{2^{4/3}}R^{1/3}=\frac{\sqrt{\alpha }}{2 \Omega_B} \Leftrightarrow \alpha =\frac{3R^{2/3}}{2^{2/3}}\Omega_B^2.
\end{equation}
Below this intersection, Two-Stream modes are directly competing with the Buneman ones. This later frontier is thus defined by
 \begin{equation}\label{eq:Bun-TS}
    \alpha=R\gamma_b^3.
 \end{equation}
The partition defined by these 3 equations in the diluted regime is thus pictured on Figure \ref{fig:partition} for $\Omega_B$ larger than 1 but smaller than $\sim R^{-1/3}$ (see Sec. \ref{sec:ultraBdiluted} for this limitation).

For larger values of $\Omega_B$, the previous trends amplify. The oblique domain (central region) is bounded towards the low $\alpha$'s because oblique growth rates are scaled like $\alpha^{1/2}$ (or $\alpha^{1/3}$), while the Buneman growth rate does not vary with $\alpha$. The evolution is of course faster with $R=1/100$ than 1/1836 because the Buneman grows faster in the second case.

From $\Omega_B\sim 5.8$, the Buneman modes gain more ``territory'' according to the aforementioned mechanisms. Noteworthily, new modes appear to govern some portion of the phase space at rather high density ratio and moderate Lorentz factor. For $R=1/1836$, the inserted spectrum maps on Fig. \ref{fig:4} show that a small part of the phase space (blue one) pertains to the oblique mode described in Section \ref{sec:ObliFinite}. This  ``middle region''  is analytically involved to explore and such results could only be derived numerically so far. The same mode governs a similar region for $R=1/100$, but in that case, another kind of oblique mode (green regions) also intervenes. Why can such modes lead the linear phase for $R=1/100$ and not $R=1/1836$? The solution lies in the mode coupling already observed in Sec. \ref{sec:2D}. Figures \ref{fig:2} (A,C) show how the Buneman modes interfere with the rest of the spectrum according to the value of $R$. For $R=1/100$, interferences are stronger than for $R=1/1836$ because the spectrum is wider. While the beam is diluted (as is the case for these plots), coupling effects are limited because the various modes are well separated in the $\mathbf{k}$ space. As the beam-to-plasma density ratio $\alpha$ increases, the Buneman spectrum progressively merges with the rest of the unstable modes. Interferences becomes thus potentially more intense, and all the more that $R$ is large. It turns out that the green region in Fig. \ref{fig:4} arises from a mode coupling which is much less excited for $R=1/1836$ than for $R=1/100$. We find here that tuning the electron-to-proton mass ratio in order to speed up the system dynamic may bear \emph{qualitative} consequences by changing the nature of the dominant mode.

\subsection{Diluted beam, ultra-magnetized regime}\label{sec:ultraBdiluted}
The outcome of the ultra-magnetized regime is relevant for Pulsars physics, for example. For diluted beams, an increasing magnetic field progressively stabilizes every modes except the Two-Stream, the Buneman and the Oblique one (see Table \ref{tab:diluted}). Both Two-Stream and Oblique are scaled like $\alpha^{1/3}/\gamma_b$, but the pre-factors favors Two-Stream modes. Competition is eventually between Buneman and Two-Stream with a frontier defined by Eq. (\ref{eq:Bun-TS}). The magnitude of $\Omega_B$ needed to trigger this picture can be derive by setting $\alpha=1$ in Eq. (\ref{eq:Ob-Bun}) as if the ``triple-point'' pictured in Fig. \ref{fig:partition} was rejected to $\alpha=1$. Such criterion places the ultra-magnetized regime beyond,
\begin{equation}\label{eq:ultra-B}
    \Omega_B\sim R^{-1/3},
\end{equation}
which fits well what is observed on Fig. \ref{fig:4}. Let us now detail the symmetric case and detail the mode hierarchy in terms of $(\gamma_b,\Omega_B)$.

\subsection{Symmetric case, $\alpha=1$}
\begin{figure*}
\begin{center}
 \includegraphics[width=\textwidth]{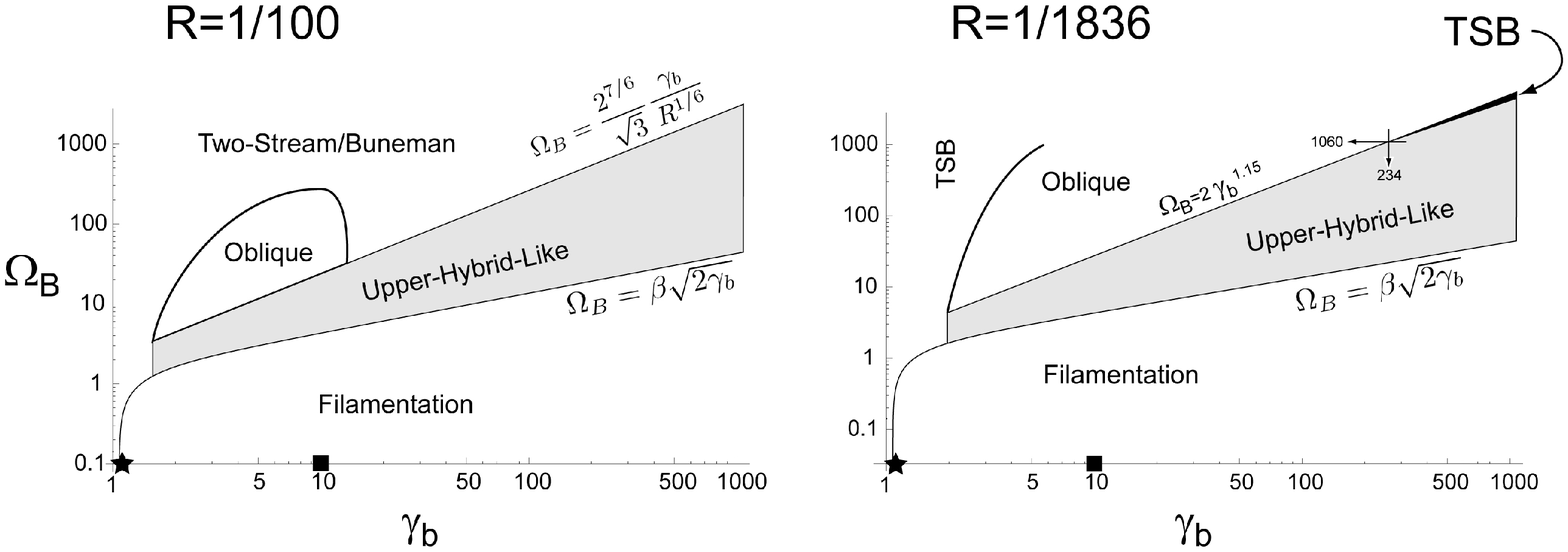}
\end{center}
\caption{Modes hierarchy for $\alpha=1$, plotted for $R=1/100$ and 1/1836. Filamentation modes (sometimes referred to as ``Weibel'') govern almost as long as their are not canceled. The black star and the square pertain to the parameters chosen for the Intergalactic Streams and Relativistic Shocks described in Secs. \ref{sec:galstreams} \& \ref{sec:shocks} respectively. The marks have been placed on the $\gamma_b$ axis because the systems they represent are un-magnetized.}
\label{fig:hierar-alf1}
\end{figure*}

When $\alpha$ reaches unity, Table \ref{tab:nondiluted} indicates that the competition is between the merged Two-Stream/Buneman, the Upper-Hybrid-Like and the Oblique modes. However, a comparison of the growth rates is possible as long as these modes are well separated in the $\mathbf{k}$ space. When overlapping, the growth rate interfere, and analytical predictions become very involved. A closer look shows that the region,
\begin{equation}\label{eq:alf1-lowB}
    \Omega_B \lesssim \sqrt{2R}\gamma_b,
\end{equation}
is thus defined. This limit lies bellow the cancelation threshold for the Filamentation instability, $\Omega_B=\beta\sqrt{2\gamma_b}$, at least up to $\gamma_b=1/R$. Indeed, the numerical exploration of this region shows that Filamentation governs the system almost as long as it is not canceled. This allows for the tracing of the Filamentation domain in  Figure \ref{fig:hierar-alf1}, together with the other limits explained in this section. As Filamentation vanishes, it evolves smoothly into the Upper-Hybrid-Like modes with $k_\parallel\propto Z_z\neq 0$.

\begin{figure*}
\begin{center}
 \includegraphics[width=0.45\textwidth]{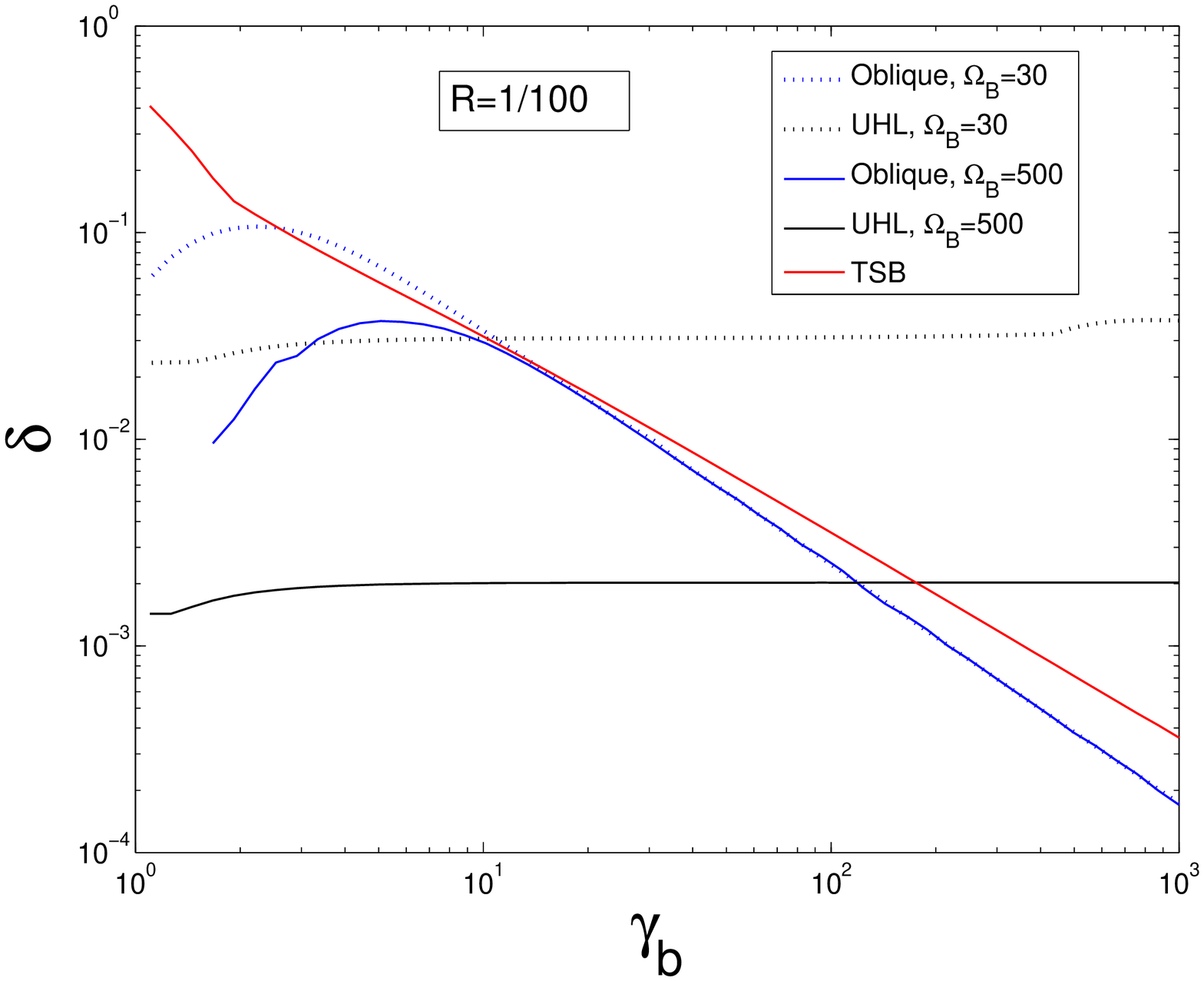} \includegraphics[width=0.45\textwidth]{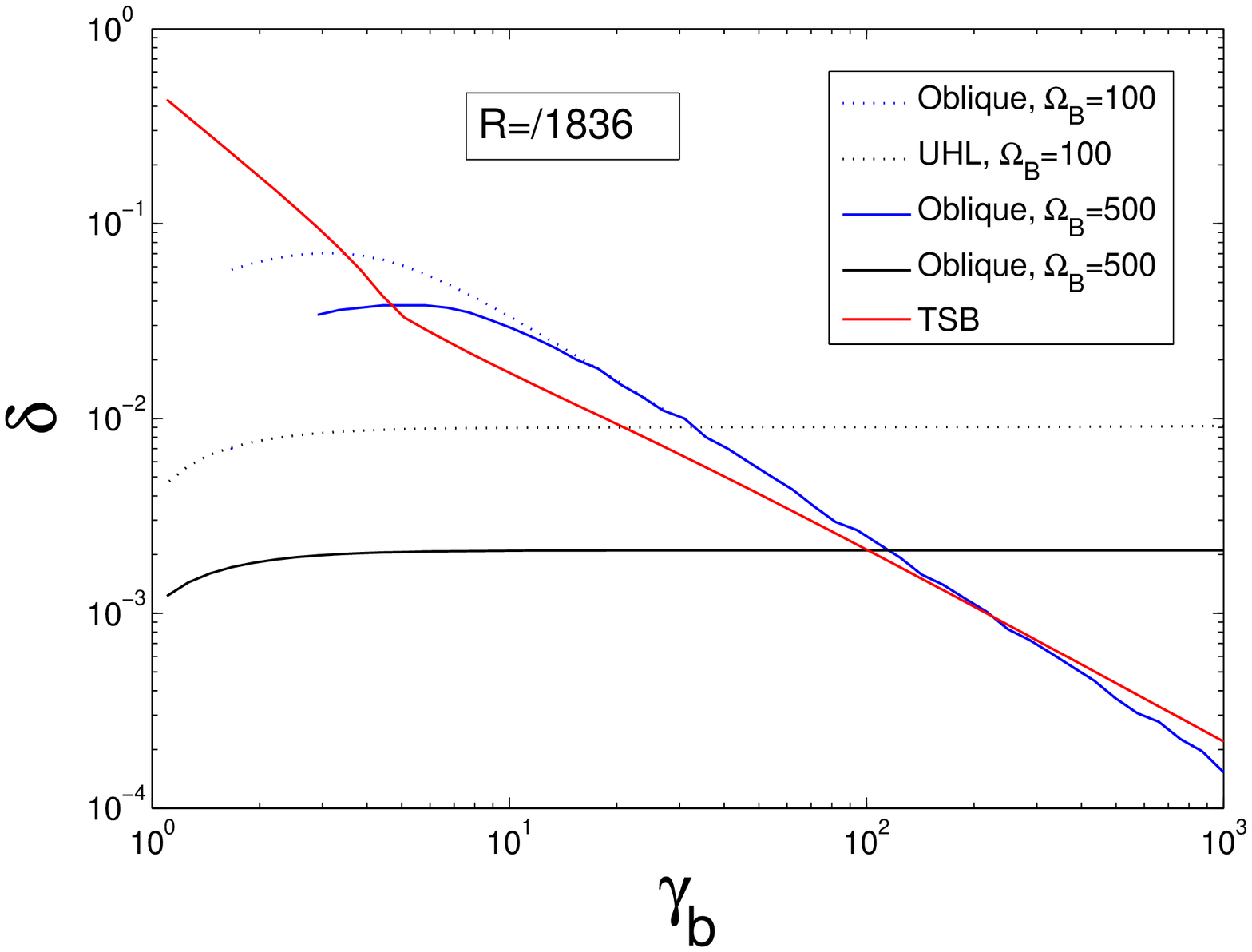}
\end{center}
\caption{(Color online) Growth rates $\delta$ ($\omega_p$ units) of the competing modes in terms of $\gamma_b$, for $\alpha=1$, $R=1/100$ and 1/1836 (numerical evaluation). The merged Two-Stream/Buneman modes (TSB, in red) varies like $\gamma_b^{-3/2}$ for $\gamma_b\lesssim R^{-1/3}$, and like $\gamma_b^{-1}$ after. Oblique modes (in blue) vary like $\gamma_b^{-1.15}$ and Upper-Hybrid-Like modes (UHL, black) like $\Omega_B^{-1}$. Due to the non-monotonic behavior of Oblique modes for moderate $\gamma_b$, the determination of the hierarchy is highly non-trivial within this energy range.}
\label{fig:compealf1}
\end{figure*}

When condition (\ref{eq:alf1-lowB}) is fulfilled, it is possible to directly compare the growth rate of the three competing modes. The resulting situation is rendered on Fig. \ref{fig:compealf1} and varies strongly with $R$. The origin of such $R$-sensitivity is the fact that only the Two-Stream/Buneman growth rate depends on the mass ratio in the large $\gamma_b$ limit. While its $\gamma_b$ variation is monotonous, the variation of the Oblique modes is not, rendering the determination of the hierarchy highly non-trivial for moderately relativistic beam.

Fig. \ref{fig:compealf1} shows that for $R=1/100$, Oblique modes are only allowed to overcome the others as long as $\Omega_B$ is not too large. Numerically, it is found that for $\Omega_B\gtrsim 300$, they are overcome by Two-Stream/Buneman regardless of the beam energy. Beyond this threshold, Upper-Hybrid-Like modes govern if
\begin{equation}\label{eq:ultra-Balf1}
    \Omega_B < \frac{2^{7/6}}{\sqrt{3}} \frac{\gamma_b}{R^{1/6}}.
\end{equation}
The resulting hierarchy map is display on Figure \ref{fig:hierar-alf1} and uncover an intriguing Bubble-like Oblique domain.

When decreasing $R$ down to its realistic value 1/1836, Fig. \ref{fig:compealf1} shows how the large $\gamma_b$ part of the Two-Stream/Buneman curve is shifted down by virtue of its $R^{1/6}/\gamma_b$ scaling (see Table \ref{tab:nondiluted}). Even for  $\Omega_B=500$, the Oblique growth rate (plain blue curve) surpasses the Two-Stream/Buneman from $\gamma_b\sim 5$ to $\sim 200$. Whether such situation holds for any $\Omega_B$ is an open question, but it has been checked that it does at least up to $\Omega_B=10^3$. In order to draw the hierarchy map, let us first determined to Lorentz factor for which Two-Stream/Buneman modes overcome the Oblique one. The corresponding value of $\gamma_b$ fulfills,
\begin{equation}
    \frac{1}{2}\frac{1}{\gamma_b^{1.15}} =\frac{\sqrt{3}}{2^{7/6}} \frac{R^{1/6}}{\gamma_b}\Rightarrow \gamma_b\sim \frac{0.055}{R^{10/9}},
\end{equation}
yielding $\gamma_b\sim 234$ for $R=1/1836$. The hierarchy map here is the one pictured on Fig. \ref{fig:hierar-alf1} for $R=1/1836$. For $\Omega_B>2\times 234^{1.15}=1060$, the system is governed with $\gamma_b$ increasing by Two-Stream/Buneman modes, then Oblique, then Two-Stream/Buneman again, then Upper-Hybrid-Like and finally, Filamentation. For smaller $\Omega_B$'s (yet, larger than $\sim 4$), the last Two-Stream/Buneman step is just skipped, as the system goes directly from the Oblique to the Upper-Hybrid-Like regime.

It is therefore uneasy to define here some ultra-magnetized regime. Unlike the diluted beam case where only two modes are left to compete for a reasonably high magnetic field defined by Eq. (\ref{eq:ultra-B}), we find here that every modes involved are likely to play a role, regardless of the magnetization. For $R=1/100$, one could defined a so-called ultra-magnetized regime from the top of the Oblique ``Bubble'' located at $\Omega_B\sim 300$. For larger magnetization, only three modes are left to compete. But such definition is no longer possible for $R=1/1836$.

Finally, let us add a remark regarding the accuracy of the borders calculations. While it is perfectly possible to numerically compute hierarchy maps \ref{fig:4} and  \ref{fig:hierar-alf1} with higher definition, such progress is not imperative because two different systems located on each side of a given border are likely to evolve in a quite similar way during the linear phase. Let us assume the border between two mode domains A and B goes through $\alpha=\alpha_0$, in such a way that A dominates for $\alpha<\alpha_0$. For $\alpha=\alpha_0$, the growth rates $\delta_A$ and $\delta_B$ are strictly equal. If $\alpha=\alpha_0\pm\varepsilon$, then $\delta_A-\delta_B=\mp\varepsilon'$.
If both A and B are excited with the same initial amplitude, the time required for the fastest to overcome the slowest only by a factor $e$ is $\sim \omega_p^{-1}/\varepsilon'$. Depending on the system considered, this time may well exceed the duration of the linear phase for $\varepsilon'$ small enough. It would thus be inappropriate to claim that only A or B are relevant on their respective side of the border. Even if the dominant mode evolution can be discontinuous, the system dynamic stemming from the growth of the entire spectrum should evolve smoothly when crossing a frontier. One can focus on either A or B only far away from the borders so that the precise calculation of its location is not crucial.

\section{Applications}\label{sec:app}
We finally turn to the determination of the modes hierarchy in terms of the parameters $\alpha$, $\gamma_b$ and $\Omega_B$ for the three settings mentioned earlier. Parameters considered in each case are reported in Table \ref{tab:param} and have been chosen after \cite{KarlickyApJ}, \cite{Lazar2009} and \cite{SilvaApJ}. The respective growth rate maps have been plotted on Figures \ref{fig:5}(A-C). Although the present theory is limited to the cold case, possible kinetic effects are indicated in each case, and even calculated for the Solar Flares environment.

\subsection{Solar Flares}\label{sec:flares}
\begin{figure}
\begin{center}
 \includegraphics[width=0.45\textwidth]{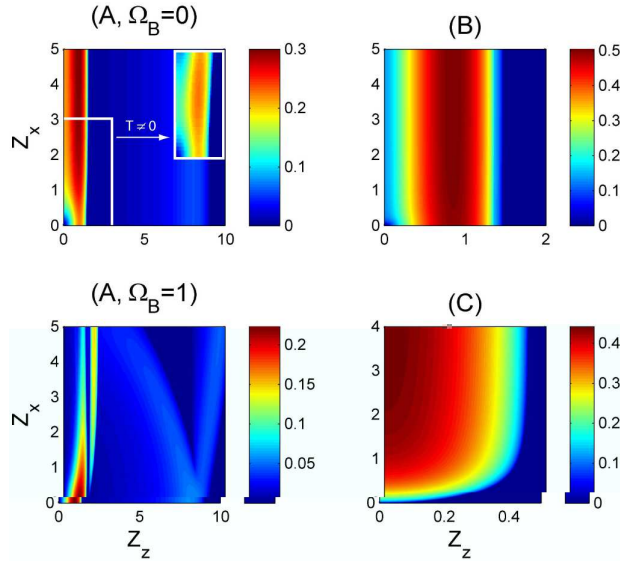}
\end{center}
\caption{(Color online) 2D unstable spectrum in terms of the reduced wave vector $\mathbf{Z}$ for parameters mentioned in Table \ref{tab:param}. (A) Solar Flares. (B) Intergalactic streams. (C) Relativistic shock. The insert on the upper left figure shows the result of a fully kinetic calculation with $R=0$, assuming electronic beam and plasma temperatures of 20 MK.}
\label{fig:5}
\end{figure}

We here perform the calculation for one magnetized and one non-magnetized case. The system location in the parameters phase space is designated by the red crosses on Fig. \ref{fig:4}, and the growth rate maps are displayed on Figs. \ref{fig:5}(A). For both scenarios, the Buneman modes are clearly visible at $Z_z\sim 8=1/\alpha$, but cannot compete with the Two-Stream ones at $Z_z\sim 1$ because the beam is not relativistic enough. With such a beam-to-plasma density ratio, a much higher relativistic factor  is required for the Two-Stream/Buneman transition.

For the non-magnetized case (upper Fig. \ref{fig:5}A), this weakly diluted beam system is governed by non-magnetized ``Oblique$_{B0}$'' modes at $Z_z\sim 1$. Although not dominant, the Filamentation instability\footnote{Denoted ``Weibel'' in \cite{KarlickyApJ}.} plays an important role as its growth rate is quite close to the largest one. The introduction of the magnetic field  (lower Fig. \ref{fig:5}A) damps Filamentation as well as oblique mode, resulting in a Two-Stream driven system. Noteworthily, Figure \ref{fig:4} shows that a small variation of the Lorentz factor can trigger a dominant mode transition for $\Omega_B=1$. For $R=1/100$, the system is close to a triple-point where Two-Stream, Buneman and Upper-Hybrid-Like modes grow the same speed. The system behavior in this case is therefore very sensitive to the parameters choice.

A word of caution is needed here before concluding with respect to the Solar Flares environment. The simulation performed in \cite{KarlickyApJ} accounts for an electronic temperatures of $21.4\times 10^6$ K (21.4 MK). Indeed, analysis of soft and hard X-ray flare data indicate that plasma
temperatures up  to  40 MK are obtained in large flares \citep{Aschwanden2002}. Furthermore, velocity distributions can be highly anisotropic due to stronger heating along the magnetic field lines \citep{Fisk1976,Miller1991,Miller1997}. For a mode with wave vector $\mathbf{k}$ growing at growth rate $\omega_i$, a temperature spread $\Delta \mathbf{v}$ can be neglected providing $|\Delta \mathbf{v}|\omega_i^{-1}\ll k^{-1}$ \citep{fainberg}. Such condition simply ensures that during one $e$-folding time, particles are traveling  almost the same path when compared to the wavelength considered. Focussing on the fastest growing Oblique$_{B0}$ mode for the non-magnetized case (upper Fig. \ref{fig:5}A), we find,
\begin{equation}\label{eq:condiT}
    \frac{|\Delta \mathbf{v}|}{c}\ll \frac{\sqrt{3}}{2^{4/3}}\left(\frac{n_b/n_p}{\gamma_b}\right)^{1/3}.
\end{equation}
Accounting for $|\Delta \mathbf{v}|=0.08c$ (i.e. $T=20$ MK), the condition above translates  $0.08\ll 0.31$. The only change when considering the Two-Stream dominated magnetized case (lower Fig. \ref{fig:5}A) is that the Lorentz factor on the right-hand side has the exponent -1 instead of -1/3. In such case, the condition reads $0.08\ll 0.25$. The cold approximation seems therefore reasonable here, although condition (\ref{eq:condiT}) is only weakly fulfilled. An interesting consequence of Eq. (\ref{eq:condiT}) is that the condition is not homogenous over the unstable spectrum. A kinetic theory of the full unstable spectrum is currently under elaboration, together with a rigorous mode-dependent definition of the kinetic-fluid transition  \citep{BretPRE2009}. For a clearer assessment of the accuracy of the cold approximation in the present case, the insert in the upper Fig. \ref{fig:5}A displays a kinetic calculation of the most relevant portion of the unstable spectrum, assuming electronic beam and plasma temperatures of 20 MK. Such calculation has been performed  neglecting ion motion, which is perfectly valid here. One can check how the portion of the unstable spectrum only slightly departs from its cold counterpart, while the largest growth rate switches from 0.3$\omega_p$ down to 0.22$\omega_p$. Note that similar calculation for the magnetized case is yet to be done.

\begin{table}
\caption{\label{tab:param}Parameters used for the Applications.}
\begin{ruledtabular}
\begin{tabular}{lllll}
Setting                        &   $n_b/n_p$   &   $\gamma_b$          &  $\Omega_B$  & Dominant mode \\
\hline
Solar Flares\footnotemark[1]   &     1/8       &  1.34    &  0, 1  &    Oblique$_{B0}$, Two-Stream          \\
Intergalactic\footnotemark[2]  &     1         &  1.005  &  0  &    Oblique$_{B0}$             \\
Shocks\footnotemark[3]         &     1         &  10                   &  0  &    Filamentation       \\
\end{tabular}
\end{ruledtabular}
\footnotetext[1]{From \cite{KarlickyApJ}, $\beta=2/3$ with $R=1/1836$.}
\footnotetext[2]{From \cite{Lazar2009}, $\beta=1/10$ with $R=1/1836$.}
\footnotetext[3]{From \cite{SilvaApJ}. Two colliding pair plasmas without background ions, i.e. $R=0$.}
\end{table}

\subsection{Intergalactic Streams}\label{sec:galstreams}
The corresponding spectrum is plotted on Fig. \ref{fig:5}B. With a small $\beta$ and a beam-to-plasma density ratio of 1, this case is also sensitive to the parameters (see the star on Fig. \ref{fig:hierar-alf1}). We find here that non-magnetized ``Oblique$_{B0}$'' modes govern the system, though Filamentation is far from being shut-down. But such a weakly relativistic symmetric system can switch between Oblique or Filamentation regimes trough very small parameters variations. When accounting for kinetic effects, Two-Stream/Buneman modes can also compete because they are the least sensitive to temperature. Indeed, for $R=0$,  these three modes grow almost exactly the same way for a beam with $\gamma_b=1.1$ and temperature 100 keV \citep{BretPRL2008}. Such result was found for a plasma temperature of 5 keV, but similar conclusions can be drawn with different plasma thermal spread. The discussion at the end of Section \ref{sec:hierar} is relevant in this case as this system is bordering various frontiers.

\subsection{Relativistic Shocks}\label{sec:shocks}
These kind of structures are currently extensively studied  by means of PIC simulations  \citep{Medvedev1999,SilvaApJ,Nishikawa2003,Milosavljevic2006,Chang2008,Spitkovsky2008,Spitkovsky2008a,Martins2009}, due to their role in the Fireball model for Gamma Ray Bursts \citep{Piran2000}. The typical scenario arising from these studies is two-fold: to start with, two relativistic plasmas shells collide and the resulting instability eventually generates a quasi-steady propagating shock. These plasma shells can be pair plasmas \citep{SilvaApJ,Chang2008}, or electron-ion plasmas \citep{Spitkovsky2008,Martins2009}. The un-magnetized pair plasma case can be analyzed here setting $R=\Omega_B=0$, and the system is found clearly governed by the Filamentation instability (see Fig. \ref{fig:5}C). This conclusion supports the emphasis put on this instability by the aforementioned authors\footnote{Here again, ``Filamentation'' is generally labeled ``Weibel'' in the literature.}. Moreover, such domination is quite robust: as long as $n_b/n_p=1$, $\gamma_b=10$ and $\Omega_B=0$, the only alternative to the Filamentation instability are the non-magnetized oblique modes. Accounting for kinetic effects, a transition to such a regime demands a beam temperature of the order $10^6$ keV \citep{BretPRL2008}. Figure \ref{fig:hierar-alf1} also suggests that some magnetized version of the system with $\Omega_B>\gamma_b\sqrt{2R}$ could trigger a transition to the magnetized oblique regime. But the bigger threat to the domination of Filamentation seems to be the beam-to-plasma density ratio. As evidenced by Fig. \ref{fig:4}, this instability dominates the ultra-relativistic regime only for strictly symmetric systems. At $\gamma_b=10$ and 100, Filamentation governs only for $n_b/n_p\gtrsim 0.6$ and 0.85 respectively. The symmetric system hypothesis has been so far related to the simulation of equivalent density colliding shells, within the Fireball framework. Since a slight change in the density ratio may trigger a dominant mode transition, subsequent work should be needed to test the robustness of the shock formation  in this respect.

The second stage of the scenario involves particles which are accelerated by the shock \citep{Spitkovsky2008a}, escape it, and interact with the upstream medium generating more instabilities. Considering now $\Omega_B=0$, $\gamma_b=100$ and $\alpha=10^{-2}$, Fig. \ref{fig:4} shows that Oblique$_{B0}$ modes dominate for $R=1/1836$ while Buneman modes do so for $R=1/100$. A more detailed evaluation of the spectrum should be done from this point, accounting for the non-thermal energy dispersion of the escaping particles, and for the upstream ion temperature. Sorting out this issue may be important because an interaction governed by the purely electrostatic Buneman instability is less likely to feed the magnetic turbulence needed in the Fireball scenario for synchrotron radiation emission. At any rate, the small beam-to-plasma density ratio implied in this second stage should prevent the Filamentation instability from playing any prominent role in the linear phase.

\section{Discussion and Conclusion}
An exact cold fluid model for a relativistic beam-plasma system including a guiding magnetic field and ion motion has been implemented, allowing for a unified description of every possible instabilities arising within such systems. After briefly reminding, or deriving, the key results for each kind of unstable modes, the hierarchy between them has been established in terms of the beam-to-plasma density ratio, the beam Lorentz factor and the magnetic field strength, considering electron-to-proton mass ratios of 1/1836 and 1/100.

In the diluted beam regime, the competition is mainly between the Two-Stream, the Buneman and the Upper-Hybrid-Like modes, as the hierarchy diagram in terms or ($\gamma_b,\alpha$) adopts a typical ``V'' shape for a given $\Omega_B$ (though not too high). Two-Stream and Buneman take advantage of the magnetic field which leave both of them unaffected while stabilizing the rest of the spectrum. Additionally, the Buneman modes, arising from the interaction of the electronic return-current with the background protons, govern the highly relativistic regime where the Two-Stream modes are strongly reduced. The ultra-magnetized regime with $\Omega_B\gtrsim R^{-1/3}$ is eventually governed by either the Two-Stream or the Buneman instabilities.

When leaving the diluted beam region, the shape of the hierarchy diagram is far from being trivial, and two different kinds of $Z_z\neq 0$ modes may dominate. The first ones are the ones already governing a part of the parameters space from $\Omega_B=0$. In the cold fluid un-magnetized model, they are found at $Z_z\sim 1$ and $Z_x=\infty$, whereas temperature effects would give them a finite perpendicular component. Under the action of the magnetic field, they continuously evolve to the Upper-Hybrid-Like modes refereed to in Table \ref{tab:diluted}, with growth rate $\propto \sqrt{\alpha}/\Omega_B$. But from $\Omega_B\sim 4.3$ and 3.6 for $R=1/1836$ and 1/100 respectively (numerical evaluations), another kind of oblique mode is likely to govern the spectrum. Unlike the previous ones, these modes reach their maximum growth rate for one single wave vector and have the growth rate scaled like $\alpha^{1/3}/\gamma_b$. With such a $\gamma_b$ scaling, their domain shrinks with increasing beam energy, as observed on Figs. \ref{fig:4}.

The transition is thus made with the symmetric regime $\alpha=1$. Here, Two-Stream and Buneman modes merge and dominate at low Lorentz factor. According to Fig. \ref{fig:hierar-alf1}, there is no such thing as a simple description of the ultra-magnetized regime. For a large mass ratio $R=1/100$ and $\Omega_B>300$, Two-Stream, Upper-Hybrid-Like and Filamentation modes are likely to dominate, depending of the Lorentz factor. At lower magnetization, Oblique modes are likely to play a role for moderate beam energy. At lower mass ratio with $R=1/1836$, the hierarchy is even more complex as the Oblique ``Bubble'' found at higher $R$ expands towards the large $\Omega_B$ regime.

While the ultra-magnetized regime is governed by electrostatic modes with flow aligned wave-vector, this is not true for density ratios close to one. Indeed, the later can be determined by virtually any kind of modes. In this respect, diluted systems are simpler to analyze, or simulate, than symmetric ones.

Some modes, like the Buneman's, strongly depends on the electron-to-proton mass ratio $R$ while others don't. As a result, the hierarchy is $R$-dependent, and lowering the ratio for computational purposes can  bring some \emph{qualitative}, not just quantitative, changes. Figure \ref{fig:4} shows for example that a system with $\alpha=0.1$, $\gamma_b=20$ and $\Omega_B=1$ is governed by Oblique modes for $R=1/1836$ and Buneman ones for $R=1/100$. Equation (\ref{eq:Bun-TS}) shows that the nature of the diluted beam ultra-magnetized regime strongly depends on $R$.

By virtue of its possible discontinuous nature, as explained in Sec. \ref{sec:hierar}, a change of the dominant mode can have dramatic consequences. One of them can be illustrated through the Two-Stream/Buneman transition defined by Eq. (\ref{eq:Bun-TS}) for diluted beams. Because Two-Stream modes are found near $Z_z=1$ and Buneman ones near $Z_z=1/\alpha$, the typical size of the structures generated  is $\omega_p/v_b$ on the left hand side of the border, and $(n_p/n_b)\omega_p/v_b$ on its right side. As long as the beam is not too diluted, these two quantities remain similar. But for a $10^3$ times diluted beam, both kind of structures differ by 3 orders of magnitude. For a relativistic symmetric system with $\alpha=1$ and $\gamma_b>\frac{1}{2}R^{-1/3}$, a transition from the Two-Stream/Buneman regime to the Upper-Hybrid-Like one implies a switch from a $Z_z=\sqrt{2R}$ dominant mode to a $Z_z=\Omega_B/\gamma_b$ one which again can span several orders of magnitude.

These  transitions have to do with the parallel wave-vector component rather than with the perpendicular one. Similar transitions have been recorded in this paper for this later component, but unstable modes located at $Z_x=\infty$ acquire a finite normal component as soon as temperature are accounted for. While kinetic effects may temper  the  $Z_x$ transitions, such should not be the case for the magnitude of the $Z_z$ transitions, because parallel components are less sensitive to temperature that normal ones.

Turning now to the astrophysical settings considered here, it is worth stressing that the value of the present theory consists more in pointing what the main instabilities \emph{could be}, rather that confirming what they actually \emph{are} (at least within the current models). Considering Solar Flares of Relativistic Shocks for example, it has already been checked that the dominant instabilities are the ones stressed in Sec. \ref{sec:app}. Even in Filamentation, or Weibel, modes are sometimes designated instead of Oblique ones, these latters are increasingly specifically discussed in the literature, and identified as such \citep{Jaroschek2005,Niemec2008,Ohira2008,frederiksenPoP2008,kong2009}. But what PIC simulations can hardly do is indicating extensively which kind of modes \emph{could} take the lead when changing the parameters, including the electron-to-proton mass ratio.

For Solar Flares, the domination of Oblique$_{B0}$ modes is quite robust in the non-magnetized case. Although the cold approximation should correctly describe the most unstable mode, given the temperatures involved, a fully kinetic calculation of the spectrum has been done to confirm this point (see upper Fig. \ref{fig:5}A). The situation is quite different when magnetizing the system. For the magnetic parameter considered here, the lower Fig. \ref{fig:5}A clearly show how some Upper-Hybrid-Like modes at $Z_z\sim 2$ can also play a role. Indeed, Fig. \ref{fig:4} shows that for a realistic value a the electron-to-proton mass ratio $R=1/1836$, the transition only needs a slightly higher relativistic factor. In such a case, the instability will heat the system both in the parallel \emph{and} perpendicular direction while the Two-Stream driven counterpart preferentially heats it along the beam direction \citep{KarlickyApJ}. Furthermore, tuning the mass ratio to speed up the dynamic can here bring qualitative consequences, as evidenced in Fig. \ref{fig:4}. It is thus found that the evolution of a magnetized electron beam in this setting, together with the kind of heating provided, is quite sensitive to the parameters.

Regarding Intergalactic Streams,  the point made in \cite{Lazar2009} is that given the parameters involved, Filamentation instability grows faster than the rest of the spectrum. The expected saturation level of the magnetic field thus produced is then found consistent with the measurements. In this respect, Fig. \ref{fig:hierar-alf1} shows that the cold version of the system is bordering the Two-Stream/Buneman-Filamentation frontier, so that Filamentation ``leadership'' is eventually  not so clear. Admittedly, kinetic effects are important as Filamentation is found to dominate through its interaction with some temperature anisotropy, but such growth rate enhancement has been found to operate in other parts of the spectrum as well \citep{BretPRE2005}. The question remains open as to know how strong Filamentation leadership needs to be for the linear phase to saturate accordingly. Such criterion would likely define some parameters window in which Weibel-like instabilities could be responsible for intergalactic magnetic fields.

Finally, the instability-based scenario for shocks formation within the Gamma Ray Bursts framework has been examined (Fireball model). Beam-plasma instabilities have been found so far to play a key role at two levels. First, they seem to prompt the shock formation itself, as the collisionless encounter of two plasma shells is unstable. Second, once the shock has been formed, it accelerates particles through diffusive Fermi acceleration. Particles escaping the shock upstream interact with the medium, and generate instabilities responsible for the magnetic turbulence needed to trigger synchronic radiation emissions.
The first unstable system has a beam-to-plasma density ratio close to unity. Indeed, it is necessarily unity in the recent PIC simulations where a beam is reflected against a wall, and eventually interact with itself. For such settings, Filamentation instability has been found leading the linear phase, and the shock formation has been so far analyze accordingly \citep{Medvedev1999,Spitkovsky2008,Spitkovsky2008a}. But the present calculations suggest that a density ratio slightly smaller than 1 may result in a quite different dominant instability. It would be very interesting to test the robustness of the shock formation scenario in this respect.
Once the shock has been formed, instabilities upstream are thus expected to generate magnetic turbulence. We now deal with a highly relativistic diluted beam/plasma system, which should definitely \emph{not} be governed by the Filamentation instability (see Fig. \ref{fig:4}). Kinetic effects are unlikely to modify this picture, as Filamentation modes are usually their first ``victims'' \citep{BretPRE2009}. Here again, a parameter window is defined allowing for the development of the best magnetic turbulence generators: Figure \ref{fig:4} clearly shows how Oblique$_{B0}$, Upper-Hybrid-Like, Two-Stream or Buneman modes are likely to shape the linear phase on this second phase. Among these four candidates, the last two are reputed \emph{purely electrostatic} instabilities, which should thus be avoided. An accurate characterization of the beam escaping the shock upstream will thus be needed to assess the validity, or set limits, to this scenario.

\acknowledgments
This work has been  achieved under projects FIS 2006-05389 of the
Spanish Ministerio de Educaci\'{o}n y Ciencia and PAI08-0182-3162 of
the Consejer\'{i}a de Educaci\'{o}n y Ciencia de la Junta de
Comunidades de Castilla-La Mancha. Thanks are due to Gustavo Wouchuk for encouraging discussions.

\appendix

\section{Tensor elements}\label{app:tensor}
The tensor mentioned in Eq. (\ref{eq:tensor}) can be cast under the from,
\begin{equation}\label{ap:T}
    \mathcal{T}=\mathcal{T}_1+\frac{1}{2x^2}\mathcal{T}_2,
\end{equation}
where
\begin{equation}\label{ap:tensors}
  \mathcal{T}_1=\left(%
\begin{array}{ccc}
  1-\frac{Z_z^2}{x^2\beta^2}    &  0                                  &   \frac{Z_xZ_z}{x^2\beta^2}       \\
  0                             &  1-\frac{Z_x^2+Z_z^2}{x^2\beta^2}   &   0       \\
  \frac{Z_xZ_z}{x^2\beta^2}     &  0                                  &  1-\frac{Z_x^2}{x^2\beta^2} \\
\end{array}
\right)
~~\mathrm{and}~~
  \mathcal{T}_2=\left(%
\begin{array}{ccc}
  T_{2xx}    &  T_{2xy}    &   T_{2xz}  \\
  T_{2xy}^*  &  T_{2yy}    &   T_{2yz}  \\
  T_{2xz}^*  &  T_{2yz}^*  &   T_{2zz}  \\
\end{array}
\right),
\end{equation}
with,
\begin{eqnarray}
  T_{2xx} &=& \frac{\alpha  \left(-x+Z_z\right)}{x \gamma_b-Z_z \gamma_b+\Omega_B}+\frac{\alpha  \left(x-Z_z\right)}{-x \gamma_b+Z_z \gamma_b+\Omega_B}-\frac{x+\alpha  Z_z}{x \gamma _p+\alpha  Z_z \gamma _p+\Omega_B}\\
  &&+\frac{x+\alpha  Z_z}{-\left(x+\alpha  Z_z\right) \gamma _p+\Omega_B}+\frac{R x (1+\alpha )}{-x+R \Omega_B}-\frac{R x (1+\alpha )}{x+R \Omega_B} ,\\
  T_{2yy} &=& \frac{\alpha  \left(-x+Z_z\right)}{x \gamma_b-Z_z \gamma_b+\Omega_B}+\frac{\alpha  \left(x-Z_z\right)}{-x \gamma_b+Z_z \gamma_b+\Omega_B}-\frac{x+\alpha  Z_z}{x \gamma _p+\alpha  Z_z \gamma _p+\Omega_B}\\
  &&+\frac{x+\alpha  Z_z}{-\left(x+\alpha  Z_z\right) \gamma _p+\Omega_B}+\frac{R x (1+\alpha )}{-x+R \Omega_B}-\frac{R x (1+\alpha )}{x+R \Omega_B}, \\
  \frac{1}{2}T_{2zz} &=& -R (1+\alpha )+x^2 \left(-\frac{\alpha }{\left(x-Z_z\right)^2 \gamma_b^3}-\frac{1}{\left(x+\alpha  Z_z\right)^2 \gamma _p^3}\right)\\
  &&+\frac{\alpha  Z_x^2 \left(\gamma_b \gamma _p \left(-\alpha  \left(x-Z_z\right)^2 \gamma_b-\left(x+\alpha  Z_z\right)^2 \gamma _p\right)+\left(\gamma_b+\alpha  \gamma _p\right) \Omega_B^2\right)}{\left(\left(x-Z_z\right)^2 \gamma_b^2-\Omega_B^2\right) \left(\left(x+\alpha  Z_z\right)^2 \gamma _p^2-\Omega_B^2\right)},\\
  T_{2xy} &=&  2 \imath R^2 x \Omega_B \frac{1+\alpha }{-x^2+R^2 \Omega_B^2}+2 \imath  \Omega_B\left(\frac{x+\alpha  Z_z}{\left(x+\alpha  Z_z\right)^2 \gamma _p^2-\Omega_B^2}+\frac{\alpha  \left(-x+Z_z\right)}{-\left(x-Z_z\right)^2 \gamma_b^2+\Omega_B^2}\right), \\
  T_{2xz} &=&  -2 \alpha  Z_x \left(\frac{\left(-x+Z_z\right) \gamma_b}{-\left(x-Z_z\right)^2 \gamma_b^2+\Omega_B^2}+\frac{\left(x+\alpha  Z_z\right) \gamma _p}{-\left(x+\alpha  Z_z\right)^2 \gamma _p^2+\Omega_B^2}\right),\\
  T_{2yz} &=&  2 \imath \alpha  Z_x \Omega_B \left(\frac{1}{\left(x+\alpha  Z_z\right)^2 \gamma _p^2-\Omega_B^2}+\frac{1}{-\left(x-Z_z\right)^2 \gamma_b^2+\Omega_B^2}\right).
\end{eqnarray}

\bibliographystyle{apj}
\bibliography{BibBret}{}

\end{document}